# Transparent metamaterials with a negative refractive index determined by spatial dispersion

*V.V. Slabko*

*Siberian Federal University, Krasnoyarsk 660041, Russia*
*Institute of Physics of the Russian Academy of Sciences, Krasnoyarsk 660036, Russia*

**Abstract:** *The paper considers an opportunity for the creation of an artificial two-component metamaterial with a negative refractive index within the radio and optical frequency band, which possesses a spatial dispersion. It is shown that there exists a spectral region where, under certain ratio of volume fraction unit occupied with components, the metamaterial appears transparent without attenuating the electromagnetic wave passing through it and possessing a phase and group velocity of the opposite sign.*

## 1. Introduction

J.B. Pendry in co-authorship papers suggested the idea of possible creation of artificial metamaterials with a negative refractive index in the gigahertz frequency band [1,2]. The idea proceeded from the work by V.G. Veselago where the processes of electromagnetic wave propagation in isotropic materials with simultaneously negative permittivity ($\varepsilon(\omega) \prec 0$) and magnetic permeability ($\mu(\omega) \prec 0$) was theoretically considered [3]. The latter leads to negative refraction index ($n \prec 0$), and phase propagation (wave vector $\vec{k}$) and energy-flow (Poynting vector $\vec{S}$) appear counter-directed, with the left-handed triplet of the vectors of magnetic field, electrical field, and the wave vector. Those were called optical negative-index materials (NIM). Interest to such metamaterials is aroused both by possible observation of extremely unusual properties, e.g. a negative refraction at the materials boundary with positive and negative refractive index, Doppler's and Vavilov-Cherenkov's effect, nonlinear wave interactions; and by possibilities in the solution of a number of practical tasks, which can overcome the resolution threshold diffracting limit in optical devices and creation of perfect shielding or "cloaking" [4-7]. At a glance, search for materials with simultaneously negative permittivity and magnetic permeability in a high frequency and, in particular, in an optical frequency band is a challenging and seemingly intractable problem. If the permittivity negative values may be obtained from materials with free charge (metals, or plasma) at the frequencies of lower plasma frequency, then materials with a negative magnetic permeability in the frequency domain over $10^{10}$ GHz are unattainable in nature. However, the design and production of artificial materials with smaller dimensions than those of the wavelength and exhibiting diamagnetic properties, is possible. On their base, metamaterials with negative refractive index may be obtained. A wide range of theoretical and experimental works done over the recent years demonstrate a feasibility of these materials in the microwave and more importantly, in the optical range of electromagnetic waves (see [8, 9]).

To study the processes of interactions between electromagnetic waves and NIM, an alternative approach producing similar results is possible. As Landau and Lifshits noted [10], the materials magnetic properties in a high frequency region and caused by a spatial nonlocality of the response at the wave magnetic field, may be described by effective permeability introduction $\tilde{\varepsilon}(\omega, \vec{k})$, which depends both on the frequency $\omega$, and on the wave vector $\vec{k}$. In this case, the magnetic materials may be treated as non-magnetic ($\mu = 1$), but the ones possessing spatial dispersion. At that, the value and direction of the wave vector is specified with $\tilde{\varepsilon}(\omega, \vec{k})$, and the energy-flow propagation is specified with group velocity $v_g = d\omega/dk$. Then, in the materials with weak absorption at a negative group velocity relation to the phase one, the Poynting vector and the wave vector are counter-directed. Here, it is relevant to address Mandelstamm's papers where it was shown that the negative refraction is a common property for the waves of any nature with negative group velocity [11]. In detail, this approach is available for investigation in the paper by Agrnovich [12]. The same author provides a thorough review of the works related to natural origin non-magnetic materials with spatial dispersion, where the appearance of effects associated with negative refractive index in a definite frequency region, is possible [13].

However, both with artificially created materials based on diamagnetic structures and with natural origin non-magnetic materials with spatial dispersion, the utilization of resonances in order to attain negative refractive indices leads to intense energy dissipation. In this case, the refractive index appears as a complex value, and the materials appear highly absorbing, which is a significant factor imposing limitations in practical applications. This paper is considering a feasibility to design metamaterials with negative refraction index determined by spatial dispersion, based on the composite of the two transparent components within the specified frequency band, with a striking difference between their refractive indices. Considered is the variant where the wave vector is real value, at negative group velocity and radiation is free of attenuation when propagated. The latter is attained through matching of the frequency and volume fraction occupied with these components.

## 2. Dispersion equation for heterogeneous materials

Let consider two-component non-magnetic materials ($\mu = 1$), where one component is a uniform medium with refractive index $n_2 = \sqrt{\varepsilon_2(\omega)}$, and which occupies a volume fractional unit $v_2$, where the particles with refractive index $n_1 = \sqrt{\varepsilon_1(\omega)}$ are uniformly placed. The volume fractional unit occupied with these particles $v_1$. We also assume that $n_1 \gg n_2$. Then,

$$v_1 + v_2 = 1 \text{ and } v_1, v_2 \succ 0 \qquad (1)$$

When the particle dimensions are small as compared to the light wavelength as in materials 2 ($\lambda_2 = \lambda_0/n_2$), and 1 ($\lambda_1 = \lambda_0/n_1$), the materials response to the field effect is spatially local, and it is easy to exhibit that one-dimensional wave equation for this materials appears as

$$\frac{\partial^2 E}{\partial z^2} + \frac{\varepsilon(\omega)}{c^2}\frac{\partial^2 E}{\partial t^2} = 0 \qquad (2)$$

Here, $\varepsilon(\omega) = v_1\varepsilon_1(\omega) + v_2\varepsilon_2(\omega)$ is effective permeability of the materials under consideration, $\lambda_0$ is electromagnetic wavelength in vacuum.

In case when particle dimensions $h$ are commensurable with wavelength λ, no consideration of this kind takes place. Indeed, the materials local polarization will be determined not only by the wave field to the point specified, but also by the field penetrating the particle with refractive index $n_1$, and reflected in various points as well. Since the particle dimensions are commensurable with the wavelength, these wave phases involve differences, and the polarization phase resultant in this point will be determined by the phases of the fields emitted in materials various points. The latter implies that the materials response in this point is local neither in time nor in space, and the spatial dispersion appears a significant factor, which determines the magnitude and direction of the wave vector and the group velocity. In this case, the permittivity $\varepsilon(\omega, \vec{k})$ exhibits the dependence upon the frequency and the wave vector.

Obviously that in the case considered, one could hardly be restricted with the approximation accepted in the description of spatial dispersion, small parameter (h/λ) expansion of dielectric susceptibility [14]. The numerical task solution for a arbitrary particle shape enables to obtain no analytic form for the dispersion law in the materials under consideration, which makes a qualitative analysis of the problem difficult. In this paper, we take up the analyses of a simple model where the approximation of geometrical optics is utilized. That enables to acquire the dispersion equation in analytical form. To describe the model, let scrutinize Fig. 1.

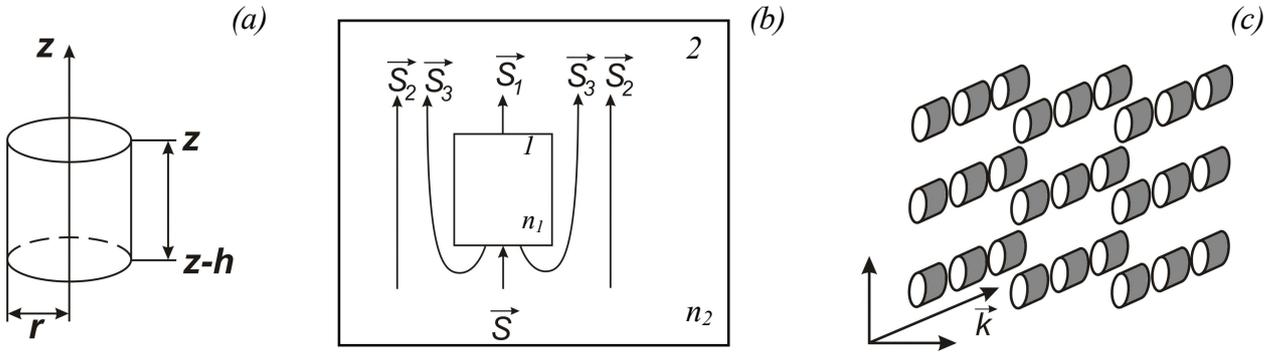

*Fig 1. (a) particle shape, (b) the scheme of the tree fields propagation in metamatherial, (c) the model of metamatherial*

Let the particles with a refractive index $n_1$, placed in the materials 2 with a refractive index $n_2$, are shaped as cylinders with a height h and the base radius r, i.e. represent a cylindrical plane-parallel plate (analog to Fabry-Perot interferometer). Here, the cylinder basis radius is assumed more than the wavelength inside the cylinder, but less than this magnitude in materials 2, and the height $h$ of order $\lambda_1 \simeq h$

$$\lambda_0/n_1 = \lambda_1 \leq d \ll \lambda_0/n_2 = \lambda_2, \qquad \lambda_1 \simeq h \qquad (3)$$

Let the wave of frequency is $\omega$ and the wave vector subject to determination $k = n(\omega, \vec{k})\omega/c$, parallel to the cylinder height, and is propagated within this materials

$$E = E_0 \exp i(\omega t - kz) + c.c. \qquad (4)$$

We assume that the energy-propagation of electromagnetic field is co-directed with the positive direction along Z-axis, and find the conditions when the field wave vector is real value, and the wave travels free of attenuation. The latter implies that in the considered field of non-absorbing components, cylinder-reflected waves appear counter-damping due to interference and the wave with a counter-directed wavevector is missing. We superpose point z and the cylinder outlet end. Then, the field at the cylinder inlet end ($z-h$) is determined by equality

$$E_h = E_0 \exp i\left[\omega t - k(z-h)\right] + c.c. \qquad (5)$$

The field on the plane coinciding with the cylinder outlet end represents the sum of the tree fields varying in phase and amplitude:
$E_1$ is the field run through the cylinder of height $h$ with refractive index $n_1$, which in turn, following the summation of the plane-parallel plate reflected from the inlet and outlet edges, looks like the Airy function [15]

$$E_1 = E\left[\left(\frac{\tau\tau' \exp i(k-k_1)h}{1-\rho'^2 \exp -i2k_1 h}\right)\right] + c.c. \qquad (6)$$

$E_2$ is the field run a distance $h$ in the materials with refractive index $n_2$

$$E_2 = E\left[\exp i(k - k_2)h\right] + c.c. \qquad (7)$$

and field $E_3$ reflected from the cylinder in plane $z=h$, enveloping it as a consequence of diffraction and propagating in materials 2, positively directed along Z-axis.

$$E_3 = E\left[\alpha\rho\left(1 - \frac{\tau\tau'\exp i(k - k_2 - 2k_1)h}{1 - \rho'^2 \exp{-i2k_1 h}}\right)\right] + c.c. \qquad (8)$$

Here $k_1 = n_1\omega/c$ и $k_2 = n_2\omega/c$ are radiation wave vectors in materials 1 and 2 subsequently, $c$ is velocity of light, $\alpha$ – is a parameter depending on the relation between the cylinder radius $r$ and height $h$, refractive indices for materials 1 and 2 contributing to the summary field, a wave $E_3$ reflected from the cylinder and diffracted in the cylinder inlet end. With respect to the above correlations, this parameter may be assumed equal to unity and is required here to follow the field effect $E_3$ on materials dispersion.

$$\tau = \frac{2n_2}{n_1 + n_2};\ \tau' = \frac{2n_1}{n_1 + n_2};\ \rho' = \frac{n_2 - n_1}{n_1 + n_2};\ \rho = \frac{n_1 - n_2}{n_1 + n_2} \qquad (9)$$

$\tau$, $\rho$, are amplitude Fresnel coefficients of transmission and reflection in transfer from material 2 to material 1 and $\tau'$, $\rho'$ from 1 to 2 subsequently. At that, $R = \rho^2 = \rho'^2$ and $\tau\tau' = 1 - \rho^2$ [15].

Let advert to wave equation (2). When the particle dimensions appear commensurable with the wavelength effective permittivity $\varepsilon(\omega) = v_1\varepsilon_1(\omega) + v_2\varepsilon_2(\omega)$, the description of the field amplitude and phase change resulting from its interactions with materials, will be transformed. Indeed, the field passing through material 1 in point z is in form (6), which is similar to introducing effective permittivity falling to one particle

$$\varepsilon_1(\omega, \vec{k}) = \varepsilon_1(\omega)\left[\left(\frac{\tau\tau'\exp i(k - k_1)h}{1 - \rho'^2 \exp{-i2k_1 h}}\right)\right] \qquad (10)$$

Then the volume unity effective permittivity determined by particles presence 1, is equal to $\tilde{\varepsilon}_1(\omega, \vec{k}) = v_1\varepsilon_1(\omega, \vec{k})$.

Similarly for the field passing through material 2 and possessing two components $E_2$ and $E_3$ (7, 8) it can be written as

$$\tilde{\varepsilon}_2(\omega, \vec{k}) = \varepsilon_2(\omega)\left[v_2 \exp i(k - k_2)h + v_1\alpha\rho\left(1 - \frac{\tau\tau'\exp i(k - k_2 - 2k_1)h}{1 - \rho'^2 \exp{-i2k_1 h}}\right)\right] \qquad (11)$$

Here the summary contribution to $\tilde{\varepsilon}_2(\omega, \vec{k})$ is determined by both material 2 of volume $v_2$ and by particles with common volume $v_1$. Then the material volume unity effective permittivity is equal

$$\tilde{\varepsilon}(\omega, \vec{k}) = \tilde{\varepsilon}_1(\omega, \vec{k}) + \tilde{\varepsilon}_2(\omega, \vec{k}) \qquad (12)$$

The expression obtained involves the dependence of permittivity both on frequency, and on wave vector. We will point out that the wave field in the materials under consideration is distinct from the harmonic field. However, for its sinusoidal component in form (4), a dispersion equation may be obtained with respect to both temporal and spatial dispersion.

If we substitute effective permittivity (12) and sinusoidal component in form (4) into wave equation (2), one finds

$$\left\{k^2 - \frac{\omega^2}{c^2}\left\{\begin{array}{l} v_2\varepsilon_2(\omega)e^{i(k-k_2)h} + \\ +v_1\left[\varepsilon_2(\omega)\alpha\rho\left(1 - \frac{\tau\tau' e^{i(k-k_2-2k_1)h}}{1-\rho'^2 e^{-i2k_1h}}\right) + \varepsilon_1\left(\frac{\tau\tau' e^{i(k-k_1)h}}{1-\rho'^2 e^{-i2k_1h}}\right)\right]\end{array}\right\}\right\}E_0 e^{i(\omega t-kz)} + c.c. = 0 \quad (13)$$

which enables to obtain a system of equations relative to two variables: desired wave vector $\vec{k}$ and relative materials volume $v_1$, occupied with particles. Indeed, adding of expression (13) in an explicit form with complex-conjugate expression, we obtain an identity in form

$$A\cos(\omega t - kz) + B\sin(\omega t - kz) \equiv 0 \qquad (14)$$

which must satisfy any temporal moment t and spatial point z, at real wave vector values with constraint $A=B=0$. Whence the equations follow

$$A = k^2 - \frac{\omega^2}{c^2}\varepsilon_2(\omega)\left\{\begin{array}{l} v_1\alpha\rho\left[\cos(k-k_2)h - \frac{\tau\tau'\left[\cos(k-k_2-2k_1)h - \rho^2\cos(k-k_2)h\right]}{1+\rho^4 - 2\rho^2\cos 2k_1 h}\right] + \\ +v_2\cos(k-k_2)h + v_1\frac{\varepsilon_1(\omega)}{\varepsilon_2(\omega)}\frac{\tau\tau'\left[\cos(k-k_1)h - \rho^2\cos(k+k_1)h\right]}{1+\rho^4 - 2\rho^2\cos 2k_1 h}\end{array}\right\} = 0 \quad (15)$$

$$B = -v_2\sin(k-k_2)h + v_1\alpha\rho\left[-\sin(k-k_2)h + \frac{\tau\tau'\left[\sin(k-2k_1-k_2)h - \rho^2\sin(k-k_2)h\right]}{1+\rho^4 - 2\rho^2\cos 2k_1 h}\right] + \\ +v_1\frac{\varepsilon_1(\omega)}{\varepsilon_2(\omega)}\frac{\tau\tau'\left[\rho^2\sin(k+k_1)h - \sin(k-k_1)h\right]}{1+\rho^4 - 2\rho^2\cos 2k_1 h} = 0 \qquad (16)$$

Equations (15), (16), conjointly with (1) ($v_1 + v_2 = 1$) form a system coupling three variables $k$, $v_1$ $u$ $v_2$. With expression of $v_2$ from (1) through $v_1$ and $v_1$, from (16) we obtain an equation for wavevector which appears the materials dispersion equation

$$k^2 - \frac{\omega^2}{c^2}\varepsilon_2(\omega)\frac{\tau\tau'\left[\sin(k_1-k_2)h + \rho^2\sin(k_1+k_2)h - \frac{\varepsilon_2(\omega)}{\varepsilon_1(\omega)}\alpha\rho\sin 2k_1 h\right]}{\left\{\begin{array}{l}\frac{\varepsilon_2(\omega)}{\varepsilon_1(\omega)}(1-\alpha\rho)(1+\rho^4 - 2\rho^2\cos 2k_1 h)\sin(k-k_2)h + \\ +\tau\tau'\left[\begin{array}{l}\frac{\varepsilon_2(\omega)}{\varepsilon_1(\omega)}\alpha\rho\left(\sin(k-2k_1-k_2)h - \rho^2\sin(k-k_2)h\right) + \\ +\rho^2\sin(k+k_1)h - \sin(k-k_1)h\end{array}\right]\end{array}\right\}} = 0 \qquad (17)$$

and an equation for the volume unity fraction occupied with particles, where the wave vector is real value and radiation is free of attenuation at propagation in the materials

$$v_1 = \frac{(1+\rho^4 - 2\rho^2 \cos 2k_1 h)\sin(k-k_2)h}{(1-\alpha\rho)(1+\rho^4 - 2\rho^2 \cos 2k_1 h)\sin(k-k_2)h +} \quad (18)$$

$$+\tau\tau'\left\{\alpha\rho\left[\sin(k-2k_1-k_2)h - \rho^2 \sin(k-k_2)h\right] + \frac{\varepsilon_1(\omega)}{\varepsilon_2(\omega)}\left(\rho^2 \sin(k+k_1)h - \sin(k-k_1)h\right)\right\}$$

The solution of these equations will provide physical interpretation under natural requirements imposed on $v_1$ (1)

$$0 \prec v_1 \prec 1 \quad (19)$$

The dependence of wave vector on frequency resulting from expression (16) represents a number of curves of quasi-periodic nature, both by frequency and by wave vector, and it possesses multiple roots at specified frequency. Fig. 2a displays a fragment of this dependence corresponding to a branch for which correlation (19) is true at $\beta = n_2/n_1$ values equal to 0.25 ($n_1$ and $n_2$ assume as non dependent on frequency). Let note that value α varying from 0 to 1, has a low effect on the dependence, and therefore is accepted as 1. Fig. 2b below represents a corresponding dependence $v_1$ on frequency utilizing correlation (18). Dimensionless frequency $X = k_1 h = \frac{\omega}{c} n_1 h$ is directed along X-axes, and dimensionless wave vector $Y = kh = \frac{\omega}{c} n(\omega,k) h$ (Fig. 2a) and (Fig. 2b.) are directed along Y-axis. Refractive index in such materials is defined by equality $n(\omega,k) = (Y/X) n_1$.

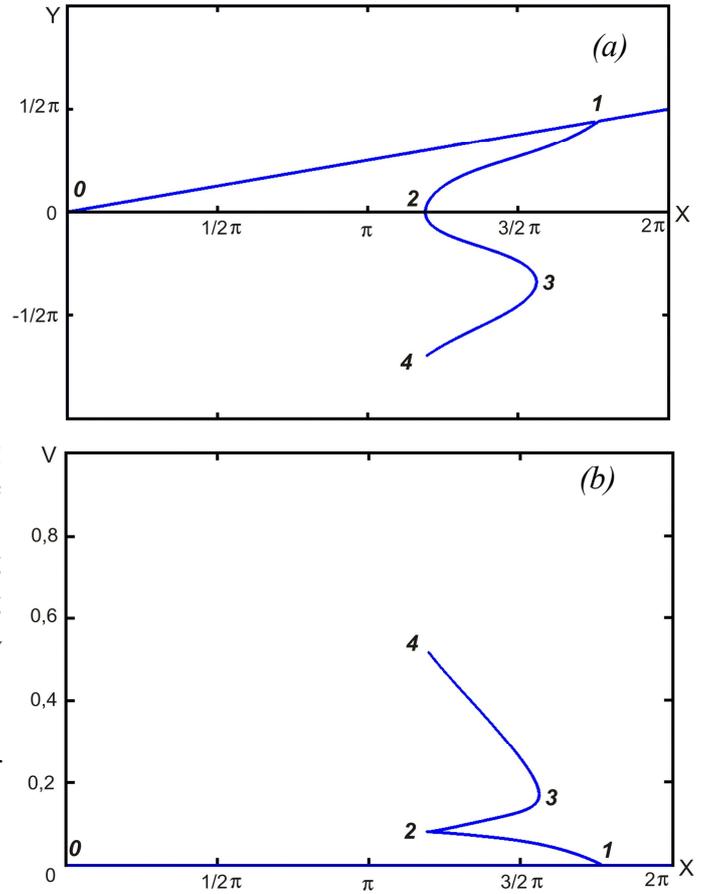

Fig 2. The frequency $X = k_1 h = (\omega/c) n_1 h$ dependence of wave vector $Y = kh = (\omega/c) n(\omega,k) h$ (a) and volume fraction $v_1$ (b). $\beta = n_2/n_1 = 0.25$

## 3. Results discussion

1. Let us point out four domains lying between characteristic points 0, 1, 2, 3, and 4 on curves 2a. and 2b. Domain 0-1 represents a trivial case when $v_1$ is equal to zero and the material refractive index is $n_2$.

In domain 1-2 the material refractive index is co-growing from 0 to $n_2$ in point 2, alongside with the frequency, which complies with gradual decreasing $v_1$ up to zero. Further the frequency increase results in negative volume fraction and the result loses physical interpretation. We note that in this domain, group velocity $v_g = d\omega/dk$ is modifying from 0 in point 2 to its extreme value in point 1, it stays positive and is co-directed with phase velocity. Zero group velocity corresponds to point 2 on the curve of dependence $v_1$ on frequency (Fig. 2b), where $dv_1/d\omega = \infty$ and is hardly attainable since it complies with strictly monochromatic radiation at exact value $v_1$ in point 2. It should be stressed that when the frequency tends to point 2, the wave vector and refractive index tend to zero, and the wavelength tends to infinity. The materials of this domain hold positive dispersion where phase and energy-flow propagation are co-directed.

The field between points 2 and 3 is inconsistent with the considered model where the energy-flow direction was initially specified as coincident with the positive direction of Z-axis. Considering that the group velocity is negative in this domain ($v_g = d\omega/dk \prec 0$), the energy-flow is co-directed with the axis negative direction, which is contrary to the model initial position.

And finally, the curve domain (Fig. 2a) starting from point 3 and being frequently lower, is characterized by negative values of wave vector $k \prec 0$ and refractive index $n(\omega,k) \prec 0$ with positive group velocity $v_g = d\omega/dk \succ 0$), which is in agreement with optical negative-index materials (NIM). In this domain $|n_2| \prec |n(\omega,k)| \prec |n_1|$ the volume fraction (Fig. 2b) corresponds to relation (19), and the group velocity is near zero in point 3. In quality, the relation exhibited in Fig. 2a and 2b stays similar to parameter $\beta = n_2/n_1 \approx 0,5$. As $\beta$ is reduced, point 2 is shifting to minor and point 3 to major frequency domain, and the frequency domain for a virtual negative dispersion is expanded. With $\beta \approx 0,5$ the position of point 2 and point 3 corresponds to similar frequency value but negative dispersion domain exists as before. We note that $n_1$ may attain values of 10 within gigahertz frequency band in a number of ceramics, and around 4 within optical band in semiconductors.

2. The accepted approximation for geometrical optics is sufficiently rough since radiation diffraction is neither allowed for inwards, nor outside the cylinder through its lateral surface, which leads to amplitude and wave-phase redistribution inside and outside of particle. Proceeding from that, it is required to give estimation defining application limits. The peculiarities covered are in agreement with the frequency close to modes $TEM_{001}$ and $TEM_{002}$ of an open resonator with plane mirrors (parameter g=1) thoroughly investigated (see e.g. [16]). Following [16], field energy losses per passage of around 10% are attained with Fresnel number $N = r^2/\lambda_1 h \approx 2$, which corresponds to $r = \lambda_1$ with $h \approx \lambda_1/2$. Field phase difference within the resonator under the same conditions will give $\Delta\varphi \approx 10^0$. Therefore, the consideration given may be of sufficient confidence with $r \geq \lambda_1$, for a qualitative analyses, at any event.

3. Modes $TEM_{001}$ and $TEM_{002}$ with the above relations between the cylinder particle radius and height, offer an extremely low mode Q factor not exceeding some unities, inasmuch as the resonant width at half-height is around the resonant frequency. With these relations, it is conceivable to attain cross-mode excitation of a very high mode Q factor, the lowest of which will be $THM_{01q}$ and $TEM_{01q}$ [17]. These mode frequencies may be obtained from frequency band $(\omega/c)n_1 h = k_1 h \leq 2\pi$ to another frequency region through appropriate relation matching between the cylinder particle radius and height. Therefore, for the covered model these resonances are negligible. Although under close examination, they may be of significant value within resonant frequency band due to their high mode Q factor.

4. The cylinder particle shape covered here is not the only one possible, and it stipulates a number of material properties utilized as the basis for its engineering. Such a material will not be critical to radiation polarization but anisotropic, because the relations obtained are co-directed with the wave vector coincident with the cylinder axis. In connection with this, investigation of a composite with sphere-shaped particles and free of the limitation mentioned, has been of our interest.

**Conclusion**

Considerations, given in this paper, exhibit the feasibility of engineering transparent metamaterials with negative refractive index in a wide frequency domain both in radiofrequency band and in optical band, on the basis of a composite with spatial dispersion. Such materials, also enable to vary the refractive index between 0 and $n_2$ and the group velocity between 0 and $v_{g2}$ in the region of positive dispersion. The engineering for such metamaterials may occur simpler than that one utilized for artificial diamagnetic structures.

*Acknowledgements.* The author is grateful to A.V. Shamshurin for accomplishment of calculations and figures.
*This work was supported by the grant of program "Development scientific potential of high school" DSP 2.1.1814, the Russian Federation President grant for scientific schools state support 3-1.3-07-12-103.*